\documentclass[twocolumn]{aastex62}
\usepackage{lineno}

\graphicspath{{./}{}}

\shortauthors{Mukherjee et al.}

\usepackage{multirow}
\begin{document}

\title{Light Curve and Hardness Tests for Millilensing in GRB 950830, GRB 090717A, and GRB 200716C}

\author{Oindabi Mukherjee}
\email{omukherj@mtu.edu}

\author{Robert J. Nemiroff}
\affiliation{Michigan Technological University\\Department of Physics\\
1400 Townsend Dr\\
Houghton, MI 49931}

\submitjournal{The Astrophysical Journal}



\begin{abstract}

To show an internal signature of gravitational lensing, two different temporal sections of a single gamma-ray burst (GRB) must be statistically similar. Here two straightforward gravitational lensing tests are defined and applied: a light-curve similarity test and a hardness similarity test. Gravitational millilensing has been claimed to be detected within several individual GRBs that contain two emission episodes separated by a time delay. However, our analyses indicate that none of those claims clearly satisfy both tests. The hardness similarity test performed on GRB 950830 and GRB 090717A found that the ratio between the second and the first emission episodes in each energy channel differed from the same ratio averaged over all detected energy channels at above the 90 percent confidence level. Also, a light curve similarity test performed on GRB 950830, GRB 090717A, and GRB 200716C separately, found that it is unlikely that the two emission episodes in each GRB were drawn from a single parent emission episode for that GRB, with differences at the 3.0 sigma, 8.3 sigma, and 8.3 sigma confidence levels respectively. 

\end{abstract}



\section{Introduction}\label{sec:intro}

If a high-enough cosmological density of dark matter (DM) compact objects (COs) were to exist in the universe \citep{1991ComAp..15..139N}, detection of gravitational lensing events would be expected in the light curves of a single Gamma-Ray Burst (GRB) \citep{paczynski1987gravitational}. If a CO happens to lie near the line of sight of a bright source, then light from that source will be bent around it due to its gravitational effects. This phenomenon is known as gravitational lensing and can be used as a tool to probe CO DM. Examples of bright sources that have been used to study such an effect are quasars and galaxies. Sources that emit huge bursts of energy in a short amount of time such as GRBs fall under the category of transient sources. In the present work, GRBs are used as probes since they occur in abundance and can be found at cosmological distances.

When COs acting as gravitational lenses have masses in the range of $10^{6}$ $M_\odot$ to $10^{9}$ $M_\odot$, the gravitational lensing effect is known as millilensing \citep{nemiroff2001limits}. The term $'milli'$ is used when the angular separation of the lensed images is on the order of milli-arcseconds, even if the images are not resolved angularly. The CO masses that can be probed using the data of individual GRBs fall in the millilensing range. The lower end of this mass range is limited by the duration and the brightness of the GRBs whereas the upper limit is decided based on the available background data following each GRB.

One of the first papers to explore this mass range was by \citet{blaes1992using}. In this paper, we explore the possibility of detecting gravitational millilensing in the light curves of GRBs. The idea of detecting COs due to gravitational lensing of distant sources was first suggested by \citet{press1973method}. Detection of gravitational lenses from any temporally-resolved sources may provide an estimate of the fraction of the universe composed of COs within a specific mass range. Alternatively, a failure to detect them will impose constraints on the values of CO mass and cosmological density.

GRBs can be used as probes to search for gravitational lensing effects as they occur in abundance and can be found at cosmological distances. 
The Burst and Transient Source Experiment (BATSE) \citep{fishman1989batse} onboard the Compton Gamma-Ray Observatory \citep{gehrels1993compton} has detected about 2700 GRBs and the currently ongoing Gamma-ray Burst Monitor (GBM) \citep{meegan2009fermi} on board the Fermi Gamma-ray Space Telescope \citep{michelson2010fermi} has detected about 3000 GRBs. Previous mililensing searches in BATSE GRBs were conducted by \citet{nemiroff2001limits}, and \citet{Marani1999GrbMillilensing}. 
\citet{ougolnikov2001search} tried to find millilensing in 1512 BATSE GRBs and \citet{li2014search} also searched for lensing in BATSE GRBs. However, none of these searches resulted in the detection of any millilensing. 

Recently, however, there have been several claims of gravitational millilensing in the light curves of single GRBs. The first claim published in the year 2021 was of GRB 950830 by \citet{paynter2021evidence}. Subsequently, two papers claiming millilensing in GRB 090717A were by \citet{kalantari2021imprints}, and \citet{2022ApJ...934..106K}. After that, two papers claiming millilensing in GRB 200716C were published by \citet{yang2021evidence}, and \citet{wang2021grb}.  All of these papers claim that the first episode of emission (hereafter called a pulse) near trigger time is gravitationally echoed by a second pulse immediately following the first pulse, thereby indicating that both pulses are gravitationally-lensed images of the same parent pulse. Even more recent claims of millilensing now exist, but they are not mentioned here because they are currently being analyzed.

There are certain criteria that need to be fulfilled to be sure that the two pulses are millilensed images of the same parent pulse and not an unrelated emission. These criteria are; 1) the emission should be significantly above the background, 2) the two pulses must have identical light curves within error margins, 3) Both the pulses should have identical spectra within error, and 4) both pulses must originate from the same location in the sky. 

Following preliminary analyses by \citet{Mukherjee_2021}, \citet{mukherjee2021light}, and \citet{mukherjee2022light} on GRBs 950830, 090717A, and 200716C, we present here more detailed analyses of the gravitational millilensing hypotheses. In comparison to the research notes, in this paper, a modified version of the light curve similarity test as detailed in section \ref{subsec:ST} is proposed. Too many bins close to the background in some GRBs might give a low chi-square value--indicating similarity although the two pulse shapes close to the peak look very different and so these results cannot be trusted. This new method would allow comparing different regions of the two pulses in a GRB, keeping the brightness factor and the time delay between the arrival of the first and the second pulse the same. The method, therefore, helps to ensure the detection of a gravitational lens. Section \ref{subsec:Data} gives a detailed description of the data used. In Sections, \ref{subsec:ST} and \ref{subsec:HT} a light curve similarity test and a hardness similarity test are described respectively. Sections \ref{subsec:950830}, \ref{subsec:090717A}, and \ref{subsec:200716C} describe the results for GRB 950830, GRB 090717A, and GRB 200716C respectively. The conclusion and summary are given in section \ref{subsec:Conclusion}.

\section{Detectors and Data}\label{subsec:Data}
The GRBs analyzed here were detected by Fermi's GBM and Compton's BATSE. The BATSE detectors were housed in 8 modules located at the corners of the CGRO spacecraft. The primary detectors in each of the modules, the Large Area Detectors (LADs), are made up of NaI (Tl) crystals that operate at an energy range of 30 keV to 1000 keV. The GBM consists of 14 detector modules; out of these 12 are Sodium Iodide (NaI) detectors and two are Bismuth Germanate (BGO) detectors. The NaI detectors cover an energy range of 4 keV to 2000 keV, whereas the BGO detectors cover an energy range of 200 keV to 40 MeV. 

The Time Tagged Events (TTE) BATSE data with a precision of 2 microseconds has been utilized here for analyzing BATSE-detected GRB 950830. The four BATSE energy channels utilized were channel 1 (20 - 60 keV), channel 2 (60 - 110 keV), channel 3 (110 - 320 keV), and channel 4 (320 - 2000 keV). 

The 64-millisecond continuous time (CTIME) GBM data and the 2-microsecond TTE data containing individual detector counts were utilized for the GBM-detected GRB 090717A and GRB 200716C. The GBM CTIME data consists of counts divided into 8 different energy channels. The energy ranges for these channels are; channel 1 (4.5 - 11.8 keV), channel 2 (11.8 - 26.9 keV), channel 3 (26.9 - 50.4 keV), channel 4 (50.4 - 101.6 keV), channel 5 (101.6 - 293.8 keV), channel 6 (293.8 - 537.8 keV), channel 7 (537.8 - 983.3 keV), and channel 8 (983.3 - 2000 keV). The GBM TTE data given in energy/photon was divided into the same 8 energy channels as the CTIME data. However, only the channels that detected an obvious signal above the background were used in the analyses of GRB 090717A and GRB 200716C.

The energy bands already defined by BATSE and the Fermi GBM are used in the hardness similarity test described in section \ref{subsec:HT}. The selection of different energy bands might affect the hardness ratio values although not likely to modify the conclusions.

\section{Light Curve Similarity Test}\label{subsec:ST}
If two pulses in a GRB are gravitational lens images of a single parent pulse, their light curves should be statistically identical. A $\chi^2$ analysis is used here to test for light-curve similarity. To compute $\chi^2$, first, the values of two independent variables were determined; one is $r_{echo}$, the ratio in fluence between the two pulses, and the other one is $t_{offset}$, the time difference between the arrival of the first pulse and the arrival of the second pulse. Specifically, $r_{echo}$ was used to artificially decrease the fluence of the first pulse in order to match it to the fluence of the second pulse, and $t_{offset}$ was used to align the two pulses temporally. Following \citep{Nemiroff2000PulseStartConjecture} and \citep{Hakkila2009GrbPulseStart}, it was assumed that $t_{offset}$ is the same at all energies and at all times. Since gravitational lensing does not change the relative timings internal to images, all source images should have the same light curve to within an amplitude scale factor $r_{echo}$. 
 
Given any $r_{echo}$, the reduced formula used for the $\chi^2$ is 
\begin{equation}
    \chi^2 =  \sum_{t}\frac{(r_{echo} P_1(t) - P_2(t))^2}{r_{echo}^2 P_1(t) + P_2(t) + B_1(t) + B_2(t)},
\end{equation} 
adapted from \citet{press1992numerical}, where
$P_1(t)$ is the number of counts in the first pulse over time,
$B_1(t)$ is the number of counts in the background underlying the first pulse over time,
$P_2(t)$ is the number of counts in the second pulse over time, and
$B_2(t)$ is the number of counts in the background underlying the second pulse over time. 
The denominator represents the total variance for $P_1(t)$ and $P_2(t)$ and therefore incorporates the Poisson noise inherent in the data. Since in the numerator, $P_1(t)$ is reduced by $r_{echo}$, the variance term for $P_1(t)$ is $r_{echo}^2$ times $P_1(t)$.  

The first step in our analysis procedure was to determine a background level for each GRB in each energy channel. This was done with a polynomial fit. Next, the counts from all the detectors and all energy channels that detected an obvious signal over the background were summed. A single background level was then found for the summed energies. Next, $t_{offset}$ and $r_{echo}$ were determined simultaneously by minimizing $\chi^2$ in a `Pulse Alignment' procedure. 

To perform pulse alignment, first, the start and end times of each pulse were found separately. The start time of the first pulse was chosen to be the time when the summed counts in bins of a specific duration increased to over some sigma ($\sigma$) value above the background fit -- nearest the pulse peak. The pulse was then considered to continue until the summed counts dropped below the same $\sigma$ value. To be consistent with a possible gravitational lens interpretation, both pulses were taken to have the same duration. Then for a range of values of $r_{echo}$, the recorded counts following the first pulse were shifted in multiples of microseconds to find the $t_{offset}$ and $r_{echo}$ that minimized $\chi^2$. 

Comparing the two pulses using a wide duration that encompasses most of each pulse may not be the best way to test the similarity between the two pulses. This is because when we compare long and faint pulse regions far from the peak (usually near the start and end of the pulses), it is likely that we will get a low value of $\chi^2$ (indicating higher chances of similarity) even though the central regions look very different. Therefore our primary light-curve similarity test involves a comparison using only bins near the pulse peak. To do this, first, the bin that contained the maximum number of counts ($peak_{bin}$) was determined for the first pulse. Then, several bins both before and after the $peak_{bin}$ were selected. This isolated a high-signal testable central region in the first pulse. Then the corresponding region of the second pulse was determined using $t_{offset}$ and the duration of the first pulse. Finally, the central regions for the two pulses were compared for light curve similarity using the $\chi^2$ formula mentioned above.

\section{Hardness Similarity Test} \label{subsec:HT}
The hardness test is based on the conjecture that the gravitational deflection of a photon does not change that photon's energy so that the gravitational lensing magnification of a source should be the same at every wavelength \citep{Paczynski1986GrbLensing}. 

For this Hardness Similarity Test, we analyze most of the measured flux for each pulse. Specifically, we use the start and end times for the first pulse as determined by a designated $\sigma$ above background, as discussed in section~\ref{subsec:ST}. We then use the consistent start and end times for the second pulse as obtained from the Pulse Alignment procedure as discussed in section~\ref{subsec:ST}, thereby ensuring that $t_{offset}$ remains the same.

Given the background fits, pulse start times, and pulse durations, the counts above the background $P_{n,c}$ were determined, where the first subscript `n' refers to the pulse number and the second subscript `c' refers to the energy channel, in both pulses, and in all energy bands. The hardness ratio for each energy channel $r_c$ is found by taking the ratio of the summed counts in the second pulse $P_{2,c}$ to the summed counts in the first pulse $P_{1,c}$ and is given by $r_c = P_{2,c} / P_{1,c}$. When summing over all energy channels, $P_{n,all}$ is also determined, followed by $r_{all}$. The errors in all of these ratios were based on the Poisson noise inherent in the backgrounds as well as the pulses \citep{Sampling1977Cochran}. 

\section{GRB 950830} \label{subsec:950830}

We analyzed the raw TTE data of BATSE which has a time resolution of 2 microseconds. This short GRB triggered three detectors: 5, 6, and 7, and is prominent in four different energy channels: 1, 2, 3, and 4. To maximize $\chi^2$ results, a time resolution of 0.004 seconds was chosen for all GRB 950830 analyses. 
 
{\it Light Curve Similarity Test:}
Due to the short duration of GRB 950830, a flat background -- a zero-degree polynomial -- was found to accurately fit the background for each of the 4 energy channels. The counts and the background fit were summed across all utilized energies. The values of $t_{offset}$ and $r_{echo}$ that minimized $\chi^2$ were found to be 0.401 seconds and 0.711 respectively. The $peak_{bin}$ for pulse 1 was found to be -0.04 seconds relative to trigger time. 

Visual inspection of the pulses showed that they differed the most nearest their peaks. We, therefore, tested several time windows around the pulse peaks to look for a maximum $\chi^2$ difference. Many time windows returned significant and similar $\chi^2$ values, but here we report on only one of them: the time window that involved 5-time bins before and 9-time bins after $peak_{bin}$. The duration of this time window was 0.06 seconds. The start time for pulse 1 was -0.060 seconds relative to the trigger time. 

\begin{figure}
    \includegraphics[width=0.9 \columnwidth, angle=0]{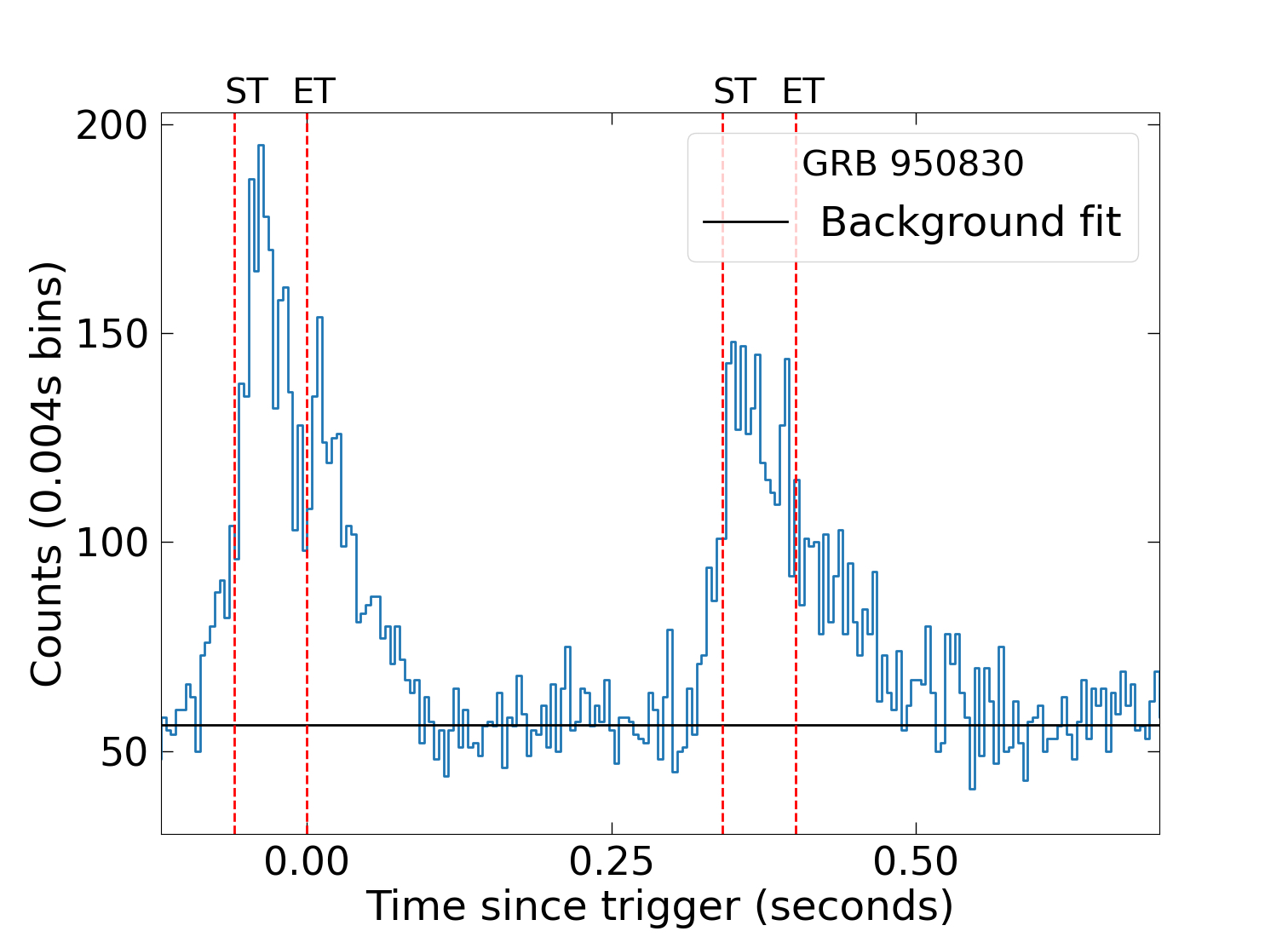}
    \caption{The light curve of GRB 950830 in 0.004-second time bins. ST and ET refer to pulse `Start Time' and `End  Time' respectively as shown by the vertical dotted lines.}
    \label{fig:950830_1}
\end{figure}

Figure~\ref{fig:950830_1} shows the regions that have been compared. A $\chi^2$ analysis found that the light curves of the two pulses differ at about the 3.0-$\sigma$ level (or 99.732 \%). 

Comparing the pulses in different time resolutions but with a similar time window affects the $\chi^2$-derived $\sigma$ value. In general, smaller bin sizes -- which correspond to higher time resolution comparisons --  resulted in a lower $\sigma$ value because both pulses are closer to the background. In contrast, larger bin sizes -- which correspond to lower time resolution comparisons -- resulted in higher $\sigma$ values. The latter trend continued up until the bin size became close to the duration of the time comparison window when significant light-curve shape information was lost. The trend remains the same for all the other GRBs that follow. Due to this inherent property of binning, the results cannot be trusted in either of the extreme situations. Hence it is necessary to check the results with several different time bins, at least up to a resolution where the shape information is intact and yet the $\chi^2$ results are not deceiving. 

{\it Hardness Similarity Test:} We first re-determined the start and end times of the first pulse from the summed counts across all energies. For this test, the start time was chosen to be the time when the summed counts in 0.004-s bins increased to over 1-$\sigma$ above the background fit -- nearest the pulse peak. The pulse was then considered to continue until the summed counts dropped to below 1-$\sigma$. The start time of Pulse 2 was found by adding $t_{offset}$ to the start time of Pulse 1. The end time of pulse 2 was chosen to be the time that maintained the same duration as pulse 1. The resulting starting times of the two pulses were determined to be -0.088 and 0.313 seconds relative to the BATSE trigger time, respectively. The duration of each pulse was found to be 0.184 seconds.

\begin{figure}
    \includegraphics[width=0.9 \columnwidth, angle=0]{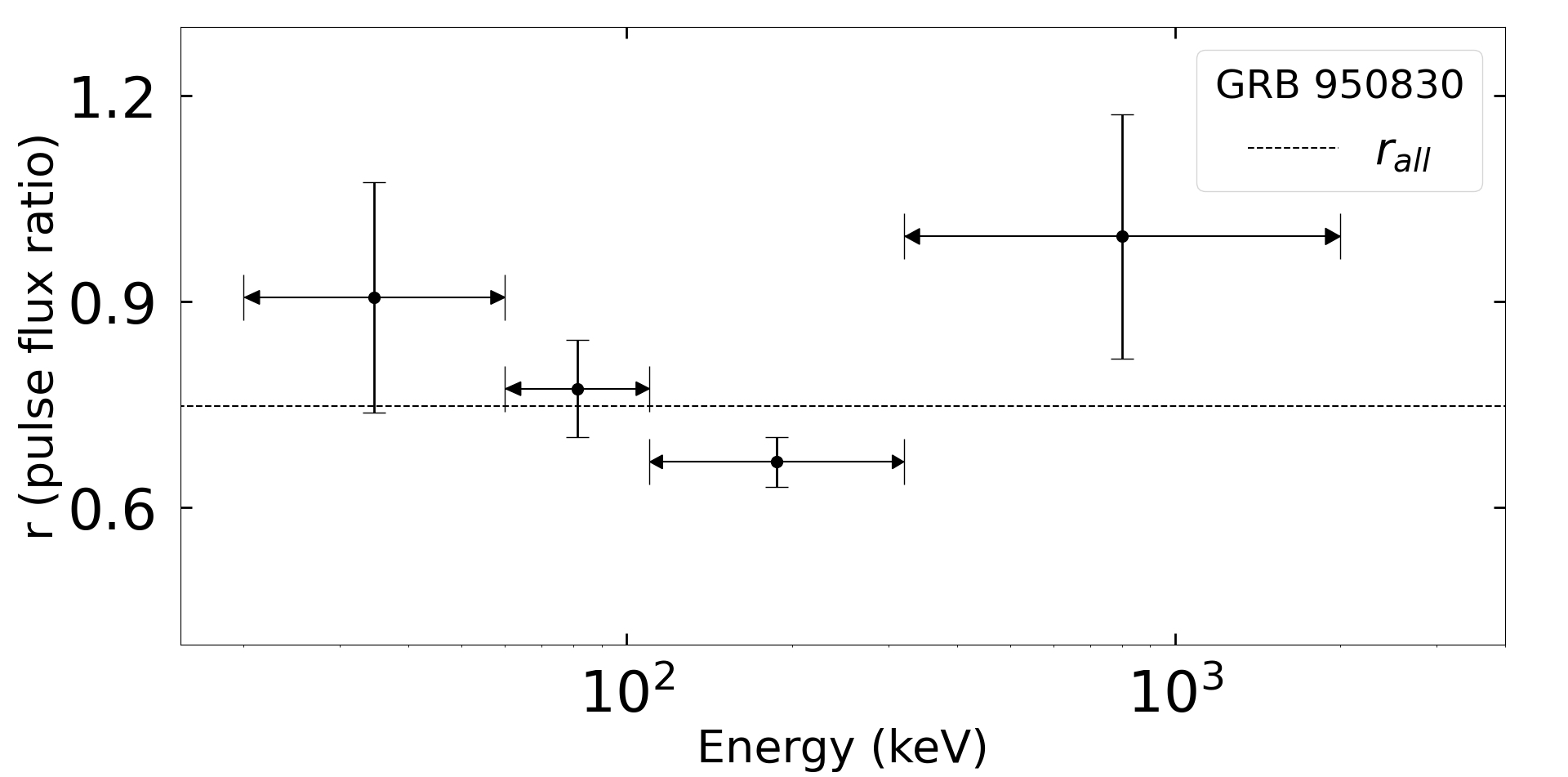}
    \caption{Count ratios between the two pulses as a function of energy channel for GRB 950830.}
    \label{fig:950830_2}
\end{figure}

As seen in Figure~\ref{fig:950830_2}, the $r_3$ value for channel 3 is somewhat lower than the other energy channels. A $\chi^2$ analysis found that the ratio $r_3$ between counts of the two pulses in BATSE energy channel 3 differs from $r_{all}$ at the 1.6-$\sigma$ level (89.289 \%).  Furthermore, channel 3 differs from channel 4 at the 1.8 $\sigma$ level (92.893 \%). From a $\chi^2$ test performed to check how different all the $r$ values were from $r_{all}$, it was found that the $r$ values differed from $r_{all}$ at about 1.9 $\sigma$ (or 93.92 \%).

\section{GRB 090717A} \label{subsec:090717A}
For GRB 090717A it was found that in Fermi GBM data, 5 of the 12 NaI detectors were pointed within 60 degrees of the burst and simultaneously detected a signal significantly above the background. These detectors were 0, 1, 2, 9, and 10. Fermi GBM energy channels 2, 3, 4, 5, and 6 were used because they recorded a prominent signal. To maximize $\chi^2$ results, a time resolution of 0.256 seconds was chosen for all analyses.

{\it Light Curve Similarity Test:} First, a third-degree polynomial was found to give an acceptable fit to the background of each of the 6 energy channels. Next, the counts and the background fit were summed across all energies. It was found that this background fit under pulse 2 did not change significantly, hence a constant background averaged over the pulse 2 region was assumed for simplicity. The $t_{offset}$ and $r_{echo}$ values that minimized $\chi^2$ were found to be 42.282 seconds and 0.542 respectively. The $peak_{bin}$ for pulse 1 was found to be 5.632 seconds. Almost all the time windows near the $peak_{bin}$ returned significant and similar $\chi^2$ values, but we will only report one of them here. The time window that involved 10 bins before and 17 bins after $peak_{bin}$ -- a total of 7.168 seconds -- gave optimal results. The resulting start time for pulse 1 was found to be 3.072 seconds relative to the trigger time. 
 
\begin{figure}
    \includegraphics[width=0.9 \columnwidth, angle=0]{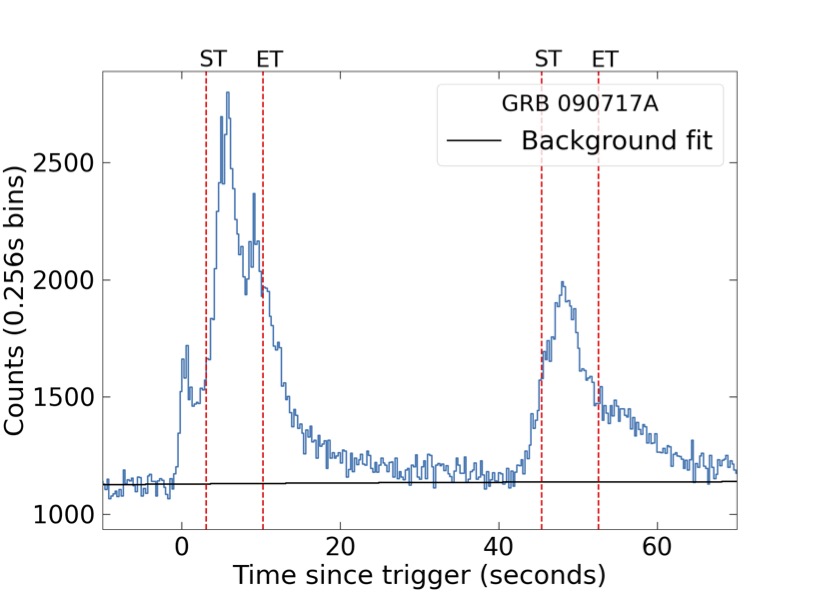}
    \caption{The light curve of GRB 090717A in 0.256-second time bins. ST and ET refer to pulse `Start Time' and `End  Time' respectively as shown by the vertical dotted lines.}
    \label{fig:090717A_1}
\end{figure}

Figure~\ref{fig:090717A_1} shows the regions that have been compared. The Light Curve Similarity test shows that the two pulses have dramatically different temporal shapes -- at above 8.3 $\sigma$ (or 99.999999999999989 \%). 

{\it Hardness Similarity Test:} For this test, the start and end times of the first pulse was found from the summed counts across all energies. The start time was chosen to be the time when the summed counts in 0.256-s bins increased to over 1-$\sigma$ above the background fit -- nearest the pulse peak. The pulse was then considered to continue until the summed counts dropped to below 1-$\sigma$. The start time of Pulse 2 was found by adding $t_{offset}$ to the start time of Pulse 1. The end time of pulse 2 was chosen to be the time that maintained the same duration as pulse 1. The resulting start times of the two pulses were determined to be -1.024 and 41.258 seconds relative to the trigger time. The duration of each pulse was found to be 22.016 seconds.

\begin{figure}
    \includegraphics[width=0.9 \columnwidth, angle=0]{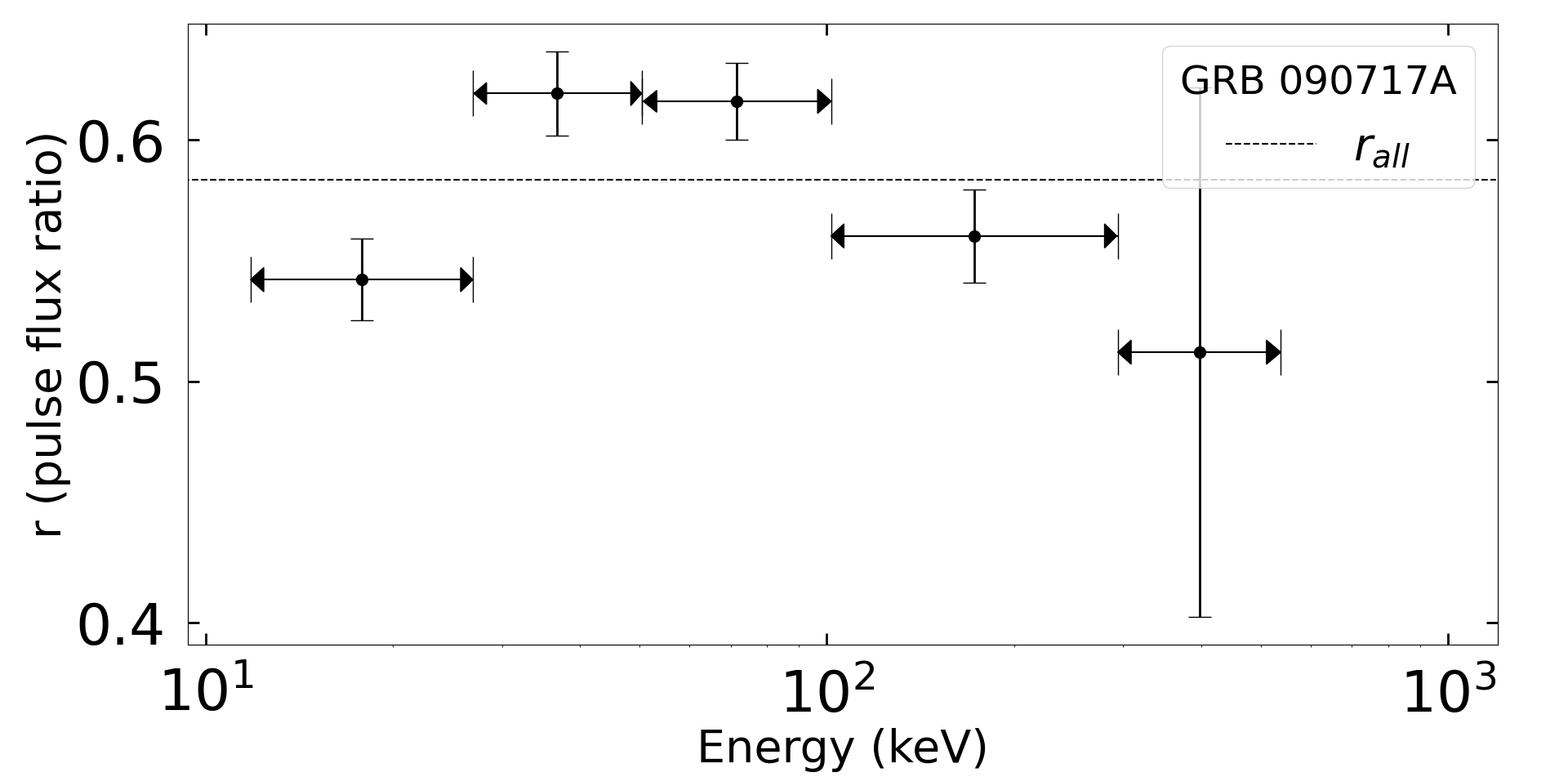}
    \caption{Count ratios between the two main pulses as a function of energy channel for GRB 090717A}
    \label{fig:090717A_2}
\end{figure}

Figure~\ref{fig:090717A_2} shows the $r$-values for the energy channels 2, 3, 4, 5, and 6, from left to right. As seen in this figure, the $r$ value for channels 2, 3, and 4, including the 1-$sigma$ error bars are different from the dotted line that depicts $r_{all}$. It was also found that the $r$ values differed from $r_{all}$ at about 1.7 $\sigma$ (or 90.953 \%). 

Some specific energy channels differed from other specific energy channels at a relatively high formal significance. For example, it was found that $r_2$ differed from $r_3$ and $r_4$ separately at the 3.2 $\sigma$ level. As these discrepancies were found later from inspection of the data, they are not definitive in excluding the lensing hypothesis.

\section{GRB 200716C} \label{subsec:200716C}
For GRB 200716C it was found that in Fermi GBM data, 4 of the 12 NaI detectors were pointed within 60 degrees of the burst and simultaneously detected a signal significantly above the background. These detectors were 0, 1, 2, and 9. Energy channels 2, 3, 4, 5, and 6 were used for all the analyses as they recorded a prominent signal. The GRB was binned up to a time resolution of 0.008 seconds to maximize $\chi^2$. 

{\it Light Curve Similarity Test:} First, a second-degree polynomial was used to fit the background of each of the 6 energy channels. Then the counts and the background fits were summed across all energies. Over the course of pulse 2, it was found that the background fit under pulse 2 did not vary significantly, hence a constant background averaged over the pulse 2 region was assumed for simplicity. The $t_{offset}$ and $r_{echo}$ values that minimized $\chi^2$ were found to be 1.909 seconds and 0.654 respectively. Next, the $peak_{bin}$ for pulse 1 was found to be 0.336 seconds relative to the trigger time. Almost all the time windows near the $peak_{bin}$ returned significant and similar $\chi^2$ values, but we will only report one of them here: the time window that involved 19 bins before and 5 bins after the $peak_{bin}$. The duration of this time window is 0.2 seconds. The pulse 1 start time was found to be 0.184 seconds relative to the trigger time. 

\begin{figure}
    \includegraphics[width=0.9 \columnwidth, angle=0]{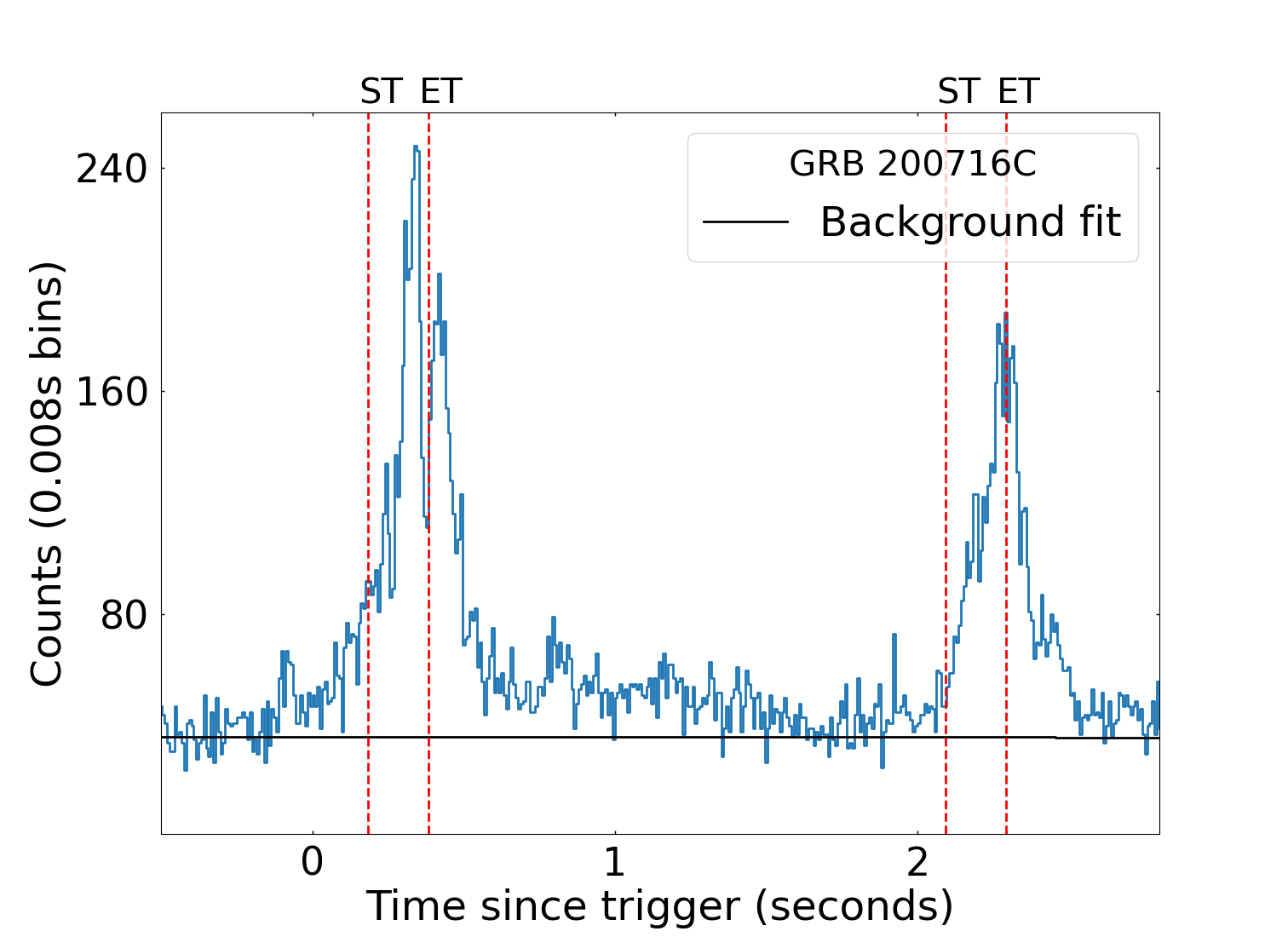}
    \caption{The light curve of GRB 200716C in 0.008-second time bins. ST and ET refer to pulse `Start Time' and `End  Time' respectively as shown by the vertical dotted lines.}
    \label{fig:200716C_1}
\end{figure}

Figure~\ref{fig:200716C_1} shows the regions that have been compared. The Light Curve similarity test shows that for a $r_{echo}$ of 0.654, the two pulses differ from each other at above 8.3 $\sigma$ confidence level (or 99.999999999999989 \%).

{\it Hardness Similarity Test:}  The start time for this test was chosen to be the time when the summed counts in 0.008-s bins increased to over 2-$\sigma$ above the background fit -- nearest the pulse peak. The pulse was then considered to continue until the summed counts dropped to below 2-$\sigma$. For this GRB, 2-$\sigma$ was used in place of the 1-$\sigma$ used for the other two GRBs because the first pulse did not sustain a return to within 1-$\sigma$ of background before the second pulse started. The start time of Pulse 2 was found by adding $t_{offset}$ to the start time of Pulse 1. The end time of pulse 2 was chosen to be the time that maintained the same duration as pulse 1. The resulting start times of the two pulses were determined to be 0.016 and 1.925 seconds relative to the trigger time. The duration of each pulse was found to be 0.56 seconds.

\begin{figure}
    \includegraphics[width=0.9 \columnwidth, angle=0]{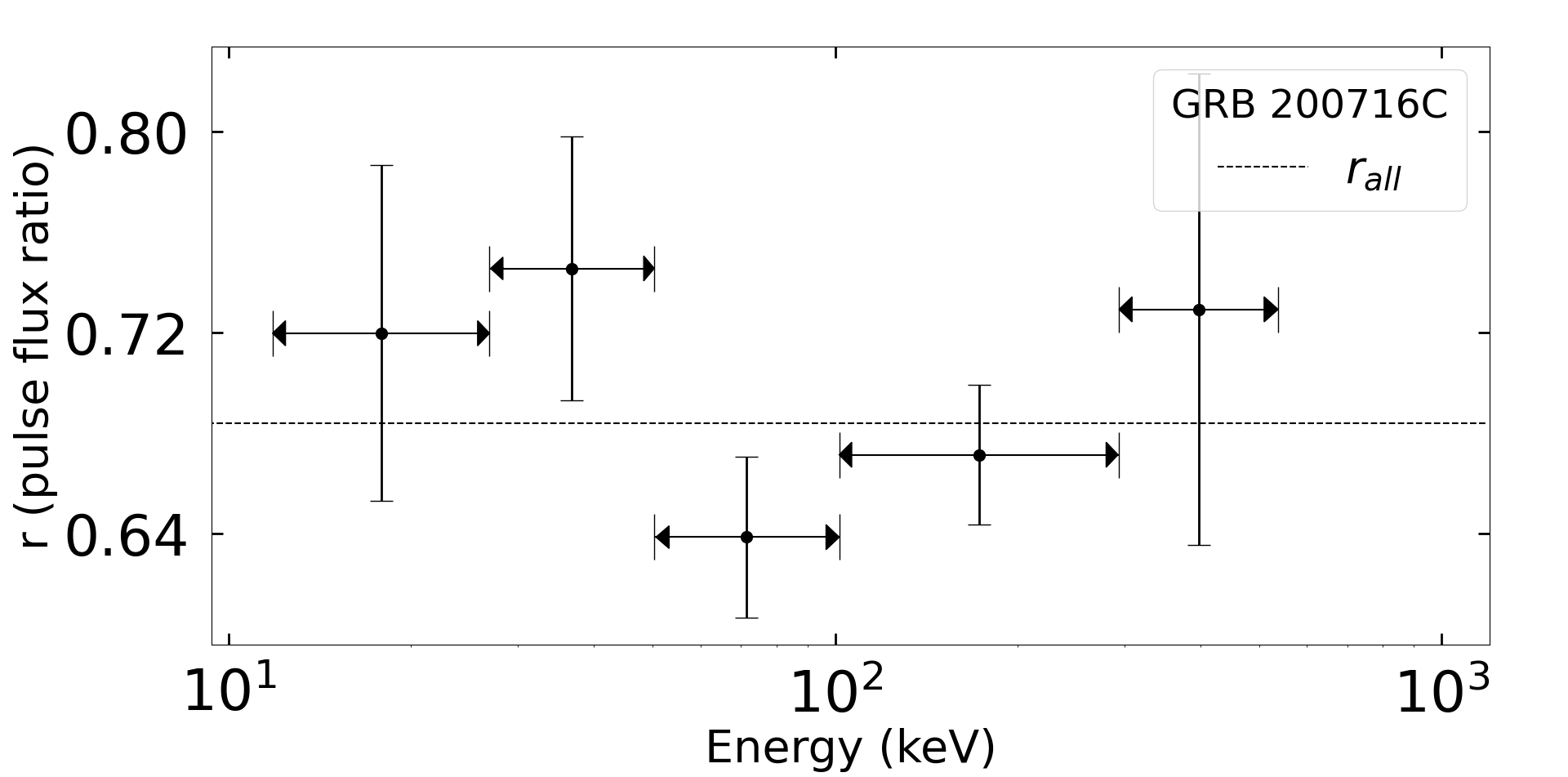}
    \caption{Count ratios between the two main pulses as a function of energy channel for GRB 200716C}
    \label{fig:200716C_2}
\end{figure}

Figure~\ref{fig:200716C_2} shows the $r$-values for the energy channels 2, 3, 4, 5, and 6, from left to right. Although it was found that the $r_2$ value differs from $r_3$ by 1.7 $\sigma$, the $r$ values of other energy channels were all in rough agreement, not only with each other but with the combined $r_{all}$. We therefore conclude that the two pulses had hardnesses that were statistically consistent.

\section{Conclusion and Summary}\label{subsec:Conclusion}
In this paper, a search for gravitational millilensing in GRB 950830, GRB 090717A, and GRB 200716C was reported. With regard to the first GRB in this list, it was found that the light curves of the two pulses in GRB 950830 differ from each other at about the 3.0-$\sigma$ level (or 99.732 \%). It was also found that the hardness ratio or $r$ values differed from $r_{all}$ at about 1.9 $\sigma$ (or 93.92 \%). 

The next burst of the three, GRB 090717A, was found to have light curves for the two bright pulses that were, at times, extremely different. Specifically, the light curves of the two pulses were found to be different at above 8.3-$\sigma$ level. In terms of hardness, the two main pulses were also found to be somewhat different: the $r$ values differed from $r_{all}$ at about 1.9 $\sigma$ (or 93.962 \%). 

Last, our analysis of GRB 200716C showed that, again, the two bright pulses were extremely different in light curve shape. Specifically, the light curves of the two pulses were found to be different at above 8.3-$\sigma$ level. The hardnesses of the two main pulses in this GRB were found to be similar -- within an expected statistical range for each to be lensed images of the same parent pulse. 

Hence we conclude that none of the three GRBs provide clear evidence in support of a gravitational lensing effect. In particular, the latter two bursts tested, GRB 090717A and GRB 200716C, even considering previously published analyses are very unlikely to involve multiple-image gravitational lensing.

We thank Michigan Technological University for their general support, and Joshua Wood for help with Fermi GBM Data Tools \citep{GbmDataTools} to find the detectors that pointed within 60 degrees of GRB 090717A and GRB 200716C separately. We sincerely thank the anonymous reviewers whose comments/suggestions helped improve the clarity of the paper.
\bibliography{aas}{}

\begin{thebibliography}{}
\expandafter\ifx\csname natexlab\endcsname\relax\def\natexlab#1{#1}\fi
\providecommand{\url}[1]{\href{#1}{#1}}

\bibitem[{Blaes \& Webster(1992)}]{blaes1992using}
Blaes, O., \& Webster, R. 1992, The Astrophysical Journal, 391, L63

\bibitem[{Cochran(1977)}]{Sampling1977Cochran}
Cochran, W.~G. 1977, Sampling Techniques, 3rd Edition. (John Wiley)

\bibitem[{Fishman {et~al.}(1989)Fishman, Meegan, Wilson, Paciesas, Parnell,
  Austin, Rehage, \& Matteson}]{fishman1989batse}
Fishman, G., Meegan, C., Wilson, R., {et~al.} 1989, in Proc. GRO Science
  Workshop, GSFC, Vol.~2

\bibitem[{Gehrels {et~al.}(1993)Gehrels, Fichtel, Fishman, Kurfess, \&
  Sch{\"o}nfelder}]{gehrels1993compton}
Gehrels, N., Fichtel, C.~E., Fishman, G.~J., Kurfess, J.~D., \&
  Sch{\"o}nfelder, V. 1993, Scientific American, 269, 68

\bibitem[{Goldstein {et~al.}(2022)Goldstein, Cleveland, \&
  Kocevski}]{GbmDataTools}
Goldstein, A., Cleveland, W.~H., \& Kocevski, D. 2022, Fermi GBM Data Tools:
  v1.1.1, , .
\newblock \url{https://fermi.gsfc.nasa.gov/ssc/data/analysis/gbm}

\bibitem[{{Hakkila} \& {Nemiroff}(2009)}]{Hakkila2009GrbPulseStart}
{Hakkila}, J., \& {Nemiroff}, R.~J. 2009, \apj, 705, 372

\bibitem[{Kalantari {et~al.}(2021)Kalantari, Ibrahim, Tabar, \&
  Rahvar}]{kalantari2021imprints}
Kalantari, Z., Ibrahim, A., Tabar, M. R.~R., \& Rahvar, S. 2021, The
  Astrophysical Journal, 922, 77

\bibitem[{{Kalantari} {et~al.}(2022){Kalantari}, {Rahvar}, \&
  {Ibrahim}}]{2022ApJ...934..106K}
{Kalantari}, Z., {Rahvar}, S., \& {Ibrahim}, A. 2022, \apj, 934, 106

\bibitem[{Li \& Li(2014)}]{li2014search}
Li, C., \& Li, L. 2014, SCIENCE CHINA Physics, Mechanics \& Astronomy, 57, 1592

\bibitem[{{Marani} {et~al.}(1999){Marani}, {Nemiroff}, {Norris}, {Hurley}, \&
  {Bonnell}}]{Marani1999GrbMillilensing}
{Marani}, G.~F., {Nemiroff}, R.~J., {Norris}, J.~P., {Hurley}, K., \&
  {Bonnell}, J.~T. 1999, \apjl, 512, L13

\bibitem[{Meegan {et~al.}(2009)Meegan, Lichti, Bhat, Bissaldi, Briggs,
  Connaughton, Diehl, Fishman, Greiner, Hoover, {et~al.}}]{meegan2009fermi}
Meegan, C., Lichti, G., Bhat, P., {et~al.} 2009, The Astrophysical Journal,
  702, 791

\bibitem[{Michelson {et~al.}(2010)Michelson, Atwood, \&
  Ritz}]{michelson2010fermi}
Michelson, P.~F., Atwood, W.~B., \& Ritz, S. 2010, Reports on Progress in
  Physics, 73, 074901

\bibitem[{Mukherjee \& Nemiroff(2021{\natexlab{a}})}]{Mukherjee_2021}
Mukherjee, O., \& Nemiroff, R.~J. 2021{\natexlab{a}}, Research Notes of the
  {AAS}, 5, 103.
\newblock \url{https://doi.org/10.3847/2515-5172/abfdbd}

\bibitem[{Mukherjee \& Nemiroff(2021{\natexlab{b}})}]{mukherjee2021light}
---. 2021{\natexlab{b}}, Research Notes of the AAS, 5, 183

\bibitem[{Mukherjee \& Nemiroff(2022)}]{mukherjee2022light}
---. 2022, Research Notes of the AAS, 6, 42

\bibitem[{{Nemiroff}(1991)}]{1991ComAp..15..139N}
{Nemiroff}, R.~J. 1991, Comments on Astrophysics, 15, 139

\bibitem[{{Nemiroff}(2000)}]{Nemiroff2000PulseStartConjecture}
---. 2000, \apj, 544, 805

\bibitem[{Nemiroff {et~al.}(2001)Nemiroff, Marani, Norris, \&
  Bonnell}]{nemiroff2001limits}
Nemiroff, R.~J., Marani, G.~F., Norris, J.~P., \& Bonnell, J.~T. 2001, Physical
  review letters, 86, 580

\bibitem[{Ougolnikov(2001)}]{ougolnikov2001search}
Ougolnikov, O. 2001, arXiv preprint astro-ph/0111215

\bibitem[{{Paczynski}(1986)}]{Paczynski1986GrbLensing}
{Paczynski}, B. 1986, \apjl, 308, L43

\bibitem[{Paczynski(1987)}]{paczynski1987gravitational}
Paczynski, B. 1987, The Astrophysical Journal, 317, L51

\bibitem[{Paynter {et~al.}(2021)Paynter, Webster, \&
  Thrane}]{paynter2021evidence}
Paynter, J., Webster, R., \& Thrane, E. 2021, Nature Astronomy, 1

\bibitem[{Press \& Gunn(1973)}]{press1973method}
Press, W.~H., \& Gunn, J.~E. 1973, The Astrophysical Journal, 185, 397

\bibitem[{Press {et~al.}(1992)Press, Teukolsky, Flannery, \&
  Vetterling}]{press1992numerical}
Press, W.~H., Teukolsky, S.~A., Flannery, B.~P., \& Vetterling, W.~T. 1992,
  Numerical recipes in Fortran 77: volume 1, volume 1 of Fortran numerical
  recipes: the art of scientific computing (Cambridge university press)

\bibitem[{Wang {et~al.}(2021)Wang, Jiang, Li, Ren, Tang, Zhou, Liang, \&
  Fan}]{wang2021grb}
Wang, Y., Jiang, L.-Y., Li, C.-K., {et~al.} 2021, The Astrophysical Journal
  Letters, 918, L34

\bibitem[{Yang {et~al.}(2021)Yang, L{\"u}, Yuan, Rice, Zhang, Zhang, \&
  Liang}]{yang2021evidence}
Yang, X., L{\"u}, H.-J., Yuan, H.-Y., {et~al.} 2021, The Astrophysical Journal
  Letters, 921, L29

\end{thebibliography}
\bibliographystyle{aasjournal}

\end{document}